\documentclass[sigconf]{acmart}

\acmConference[ICSE 2024]{46th International Conference on Software Engineering}{April 2024}{Lisbon, Portugal}

\usepackage{graphicx} 
\usepackage{subfig}
\usepackage{xcolor}
\usepackage[inline]{enumitem}
\usepackage{mdframed}
\usepackage{comment}
\usepackage{booktabs}
\usepackage{float}
\usepackage{multirow}
\setlist[enumerate,1]{label=\textit{\alph*)}}
\usepackage{amsmath,amsfonts}
\usepackage{textcomp}
\usepackage{hyperref}
\usepackage{flushend}
\usepackage{todonotes}

\newcommand{\Rbb}{\mathbb{R}}

\newcommand{\A}{\mathbf{A}}
\newcommand{\B}{\mathbf{B}}
\newcommand{\Cbf}{\mathbf{C}}

\newcommand{\F}{\mathbf{F}}
\newcommand{\Ft}{\tilde{\F}}

\newcommand{\Y}{\mathbf{Y}}
\newcommand{\Z}{\mathbf{Z}}

\RequirePackage[american]{babel}
\addto\captionsamerican{}

\usepackage[most]{tcolorbox}
\NewTColorBox{custombox}{o}{
    colback=gray!10,
	colframe=gray!10,
	left=1.0mm,
	right=1.0mm,
	top=1.0mm,
	bottom=1.0mm,
	fonttitle=\bfseries,
	arc=0mm,
	leftrule=0mm,
	rightrule=0mm,
	toprule=0mm,
	bottomrule=0mm,
	notitle,
	before=\par\medskip\noindent,
    IfValueTF={#1}{before upper={\textbf{#1: } }}{},
    parbox=false,
}

\copyrightyear{2024}
\acmYear{2024}
\setcopyright{acmlicensed}\acmConference[ICSE '24]{2024 IEEE/ACM 46th International Conference on Software Engineering}{April 14--20, 2024}{Lisbon, Portugal}
\acmBooktitle{2024 IEEE/ACM 46th International Conference on Software Engineering (ICSE '24), April 14--20, 2024, Lisbon, Portugal}
\acmPrice{15.00}
\acmDOI{10.1145/3597503.3623314}
\acmISBN{979-8-4007-0217-4/24/04}

\begin{document}

\title{Towards Reliable AI: Adequacy Metrics for Ensuring the Quality of System-level Testing of Autonomous Vehicles}

\author{Neelofar Neelofar}
\affiliation{%
  \institution{Monash University}
  \country{Australia}
  }
\email{neelofar.neelofar@monash.edu}

\author{Aldeida Aleti}
\affiliation{%
  \institution{Monash University}
  \country{Australia}
  }
\email{aldeida.aleti@monash.edu}

\begin{abstract}\label{abstract}
AI-powered systems have gained widespread popularity in various domains, including Autonomous Vehicles (AVs). However, ensuring their reliability and safety is challenging due to their complex nature. Conventional test adequacy metrics, designed to evaluate the effectiveness of traditional software testing, are often insufficient or impractical for these systems. White-box metrics, which are specifically designed for these systems, leverage neuron coverage information. These coverage metrics necessitate access to the underlying AI model and training data, which may not always be available. Furthermore, the existing adequacy metrics exhibit weak correlations with the ability to detect faults in the generated test suite, creating a gap that we aim to bridge in this study.

In this paper, we introduce a set of black-box test adequacy metrics called "Test suite Instance Space Adequacy" (TISA) metrics, which can be used to gauge the effectiveness of a test suite. The TISA metrics offer a way to assess both the diversity and coverage of the test suite and the range of bugs detected during testing. Additionally, we introduce a framework that permits testers to visualise the diversity and coverage of the test suite in a two-dimensional space, facilitating the identification of areas that require improvement.

We evaluate the efficacy of the TISA metrics by examining their correlation with the number of bugs detected in system-level simulation testing of AVs. A strong correlation, coupled with the short computation time, indicates their effectiveness and efficiency in estimating the adequacy of testing AVs.
 
\end{abstract}

\maketitle

\section{Introduction}\label{sec:introduction}

Thanks to the availability of big data and computational power, AI-based systems, like AVs, have shown a remarkable increase in their capabilities. However, performing quality assurance of these systems requires a paradigm shift, as not all the logic is explicitly encoded in the source code. These systems are trained on enormous quantities of data and act based on the logic defined in terms of patterns observed in the data~\cite{manning2008introduction}. Therefore, traditional test adequacy measures can't be effectively applied to assess the quality and reliability of these systems, and novel contributions from researchers and practitioners are urgently required. 

The challenge of testing AI-based systems is well accepted by the Software Engineering community, which is actively working on finding effective test adequacy metrics for such systems~\cite{zhang2021fairness, zhang2020machine, riccio2020testing, sun2021coverage, martinez2022software}. Since traditional white box code coverage metrics can not be effectively used to measure how well the program logic -- that is partially determined by the underlying training data -- has been adequately exercised, specialised white box adequacy metrics aiming at maximising neuron coverage~\cite{guo2018dlfuzz, pei2017deepxplore, tian2018deeptest, xie2019deephunter, kim2019guiding} or surprised coverage~\cite{kim2019guiding} are introduced for model-level testing. A limitation of the coverage-based adequacy measures is that it requires full access to the underlying DNN and training data, both of which are not often available to testers when performing system-level testing. Furthermore, the performance of most coverage measures is assessed using adversarial inputs, thus focussing on the robustness of the model instead of correctness. Irrespective of the claimed sensitivity of these measures to adversarial inputs, studies have failed to find a significant correlation between these coverage measures and their fault detection capability~\cite{aghababaeyan2021black, li2019structural, chen2020deep}. 

Another widely used adequacy measure for traditional and AI-based systems is test suite diversity computed on test inputs or outputs~\cite{aghababaeyan2021black,feldt2016test, lu2021search,birchler2021automated}. The diversity metrics are designed based on the intuition that similar test cases exercise similar parts of the source code or training examples, thus revealing the same faults. On the other hand, a system that is tested under a wide range of conditions is more likely to perform reliably in the real world. Therefore, it is believed that diversifying test scenarios improves the exploration of the fault space, leading to increased fault detection~\cite{cartaxo2011use,aghababaeyan2021black,zohdinasab2021deephyperion}. However, the current research on testing AI-based systems has identified a significant gap in the availability of good diversity metrics that show a strong correlation with fault detection~\cite{aghababaeyan2021black}. 

Among many applications of AI in Software Engineering, AVs are one of the most sophisticated and highly complex systems that rely on a combination of sensors to collect data, AI-based systems to process the data and take decisions, and control systems to operate the AV safely in diverse real-world conditions. The diversity in the testing of AVs is crucial to maximising the probability of testing the system under boundary conditions or rare emergency situations that may occur in the real world. Furthermore, AVs interact with other road users, including pedestrians, cyclists and other vehicles driven by humans. Testing under diverse conditions is important to find how the system handles potentially unpredictable and irrational actions related to human behaviour. In addition, evaluating the coverage of testing is crucial to determine the extent to which the input/output behaviour of the AV system has been tested.

Given the significance of both diversity and coverage for black-box testing of AVs, we propose a novel set of metrics, called \textit{Test suite Instance Space Adequacy (TISA)} metrics, that provides an objective measure of the quality of testing in terms of both diversity and coverage. TISA metrics are computed by identifying the features of the test scenarios having maximum impact on test outcome (safe vs unsafe), and then projecting these test scenarios -- defined in terms of the impactful features -- on $2D$ space, using a framework called Instance Space Analysis (ISA)~\cite{smith2022instance}. The representation of the test scenarios in a 2D space facilitates a clear visualisation of the test suite, resulting in an easy evaluation of the diversity and coverage of the whole test suite across each feature, along with the diversity of the detected bugs. As a result, the proposed metrics serve as a valuable tool for testers to identify areas where the test suite may lack diversity or areas that require additional testing. This ultimately leads to an enhanced quality of testing by highlighting specific areas that need improvement or further exploration.

To investigate the effectiveness of the proposed test adequacy metrics, we analyze their correlation with the number of identified bugs. Our findings reveal a strong correlation between the TISA metrics and bug count. In contrast, we observe zero, weak, or inconsistent correlations with other test diversity metrics studied. This indicates that the TISA metrics provide a unique perspective on system diversity, complementing existing methods. Additionally, their low computational cost compared to alternative approaches further justifies their suitability for assessing large test suites' adequacy.

Overall, the main contributions of this study are as follows:

\begin{itemize}
    \item We comprehensively investigate the state-of-the-art diversity metrics for testing AI-based systems and assess their effectiveness in terms of fault detection.
    \item We propose a set of novel metrics, called \textit{TISA}, which provide visual insights into the diversity and coverage of test scenarios along with providing an objective measure of the quality of testing. We assess the effectiveness of the proposed metrics for their correlation with the fault detection capability of the test suites.
    \item We analyze how TISA metrics correlate with state-of-the-art diversity metrics. 
    \item We assess the performance of TISA and other selected metrics in terms of computation time.
    \item A comprehensive replication package to reproduce the results obtained in this study is made available at \url{https://doi.org/10.5281/zenodo.7784015}.
\end{itemize}

\section{Black-box Testing Adequacy Measures} \label{sec:ISA}
For this study, we select widely used adequacy metrics for black-box system testing of AI-based systems. The available coverage based metrics are either white-box~\cite{pei2017deepxplore, ma2018deepgauge, sun2019structural, kim2019guiding, gerasimou2020importance}, or task specific~\cite{hauer2019did, arcaini2021targeting, tang2021collision, laurent2022parameter}, hence do not fit well for this study. Therefore, all the metrics we selected are diversity-based adequacy measures. 

\subsection{Shannon's Diversity Index} 
Shannon's Diversity Index is a widely accepted measure of diversity in Ecology, where diversity is defined as ``the variety and abundance of species in a defined unit of study''~\cite{magurran2021measuring}. It measures the diversity of a sample in terms of richness and evenness. \textit{Richness} measures the number of species present in a population, while \textit{evenness} measures the number of individuals per specie~\cite{magurran2021measuring}.

This concept of diversity has been adapted in software engineering, where a dataset is considered diverse, over a particular attribute, if it contains a rich variety of values of the attribute, and those values are evenly abundant~\cite{pham2010visualization}. It has been used as a measure of diversity in software testing and defect prediction studies~\cite{bennin2017mahakil, bohme2018stads}. 

The Shannon's Diversity Index is mathematically calculated using equation~\ref{eq:shannon_index}:
\begin{equation}
\label{eq:shannon_index}
    div_{Shannon}(X) = -\sum_{i=1}^s p(x_i) \ln p(x_i)
\end{equation}
where $X$ is the attribute under consideration, $\{x_1, x_2 \cdots x_n\}$ is the set of possible values of $X$, $p(x_i)$ is the probability that $X$ will take the value $x_i$. The probability $p(x_i)$ is the ratio of the number $n_i$ of instances having $x_i$ value for the attribute, to the total number $N$ of individuals in the set, i.e. $p(x_i) = \frac{n_i}{N}$.

$div_{Shannon}(X)$ is directly proportional to the level of diversity within a single attribute; a higher value indicates a richer variety and even distribution.  However, comparing $div_{Shannon}(X)$ values across attributes can be challenging as it is scaled to the number of possible attribute values, which can create discrepancies between attributes with varying levels of possible values. To address this, an \textit{evenness measure} has been adapted to normalise the value of $div_{Shannon}(X)$ by its maximum possible value, ensuring that all attributes are considered equal~\cite{bohme2018stads}. We utilise this normalised Shannon's Index in our analysis. 

\subsection{Euclidean Diversity}
Euclidean diversity ($div_{Eu}$) is the measure of the distance between two test instances in an Euclidean space. This measure of diversity has widely been used as an adequacy metric in the testing of AVs~\cite{birchler2022single, birchler2021automated, lu2021search, kim2019test, lu2022learning}. \color{black}We calculate $div_{Eu}$ following the approach adopted in~\cite{lu2021search}, which is summarised below. 

As different features of test scenarios are represented by a wide range of values, the feature values are first scaled between $[0, 1]$ using Min-Max normalisation~\cite{saranya2013study}. Next the diversity of the $k^{th}$ feature of two scenarios $S_i$, $S_j$ is calculated as below:
\begin{equation}
    div_{feat_k} = nor(|f_{ik} - f_{jk}|)
\end{equation}
where $f_{ik}$ and $f_{jk}$ are the $k_{th}$ feature values of $S_i$ and $S_j$ respectively. The diversity of $S_i$ and $S_j$ is then calculated by aggregating the diversities of all the features of these scenarios. 
\begin{equation}
    div_{i,j} =  \frac{\sum_{k=0}^{m}div_{feat_k}}{m}
\end{equation}
where $m$ is the total number of features used to define the scenarios. The diversity of a scenario $S_i$ is then calculated as:
\begin{equation}
    div_S = \sum_{j=0}^{ns}div_{i,j}, j\neq i 
\end{equation}
where $ns$ is the total number of scenarios. 

\subsection{Normalised Compression Distance}
Normalised Compression Distance (NCD) is a feature-free, parameter-free, and alignment-free similarity metric based on compression~\cite{cohen2014normalized}, and has found many applications in clustering, classification, pattern recognition, software testing, and phylogeny~\cite{kirk2004information, ane2005missing, kocsor2006application, kulesza2012determinantal, henard2016comparing, aghababaeyan2021black, coltuc2018use, cilibrasi2005clustering}. The metric is based on \textit{Kolmogorov complexity}~\cite{kolmogorov1965three} and \textit{Information distance}~\cite{bennett1998information}. \textit{Kolmogorov complexity} defines the information in a single object, while \textit{Information distance}~\cite{bennett1998information} measures the information required to transform one object into the other, among a \textit{pair} of objects. 

Cohen et al. extended NCD to support its application for calculating the similarity of  multisets~\cite{cohen2014normalized}. Due to the complexity of Kolmogorov complexity, they approximate it through real-world compressors~\cite{li2004similarity}. However, the choice of the compressor is important as different compressors would impact the performance of NCD in terms of computation time, compression distance and used memory~\cite{feldt2016test}. We followed Aghababaeyan et al. and used \textit{Bzip2} that was found to be the best in their experiments for computing NCD for black-box testing of DNNs~\cite{aghababaeyan2021black}. 

For a multiset $S$, $NCD$ metric is calculated through an intermediate measure $NCD_1$~\cite{cohen2014normalized, aghababaeyan2021black}:
\begin{equation}
    NCD_1(S) = \frac{C(S) - min_{s \in S}\{C(S)\}}{max_{s \in S}\{C(S) \setminus \{s\}\}}
\end{equation}
\begin{equation}
    NCD(S) = \max \left \{  NCD_1(S), max_{Y \subset S}\{NCD(Y)\}  \right\}
\end{equation}
where C(S) is the length of the multiset ($S$) after compression. 
One limitation of $NCD$ is the high computational cost that makes it impractical for big datasets.  

\subsection{Standard Deviation}
Standard deviation, $STD$, measures the variability or dispersion of data around the mean value and is widely recognised as a robust statistical measure to quantify the variability or spread of a test suite~\cite{aghababaeyan2021black, cai2018diversity, nikolik2006test}. For calculating $STD$ for this work, the feature values are first scaled between $[0, 1]$ using Min-Max normalisation~\cite{saranya2013study}, and $STD$ is then computed as the norm of the $STD$ of each feature.

\section{Test suite Instance Space Adequacy Metrics} \label{sec:ISA}

Test suite Instance Space Adequacy (TISA) metrics measure the test suite adequacy from a coverage and diversity perspective. Along with providing an objective measure of the adequacy of a test suite, these metrics offer valuable insights into the relationship between the structural properties of the test instances and their impact on test outcomes. To estimate TISA metrics, we employ a framework called Instance Space Analysis (ISA)~\cite{smith2022instance} that projects test instances -- characterised in terms of features -- from $n$-dimensional feature space to a $2$-dimensional space, called \textit{Instance Space} (IS). The projections are performed in such a way that there is a clear distinction between failing and passing test scenarios, and the impact of each feature on the test outcome (fail or pass) can easily be mapped. In the context of AV testing, a test scenario is deemed effective if it reveals a bug in the system. As we are using black-box system testing of AVs as a case study, an effective scenario is one that reveals a failure of the AV. Depending on the particular module of the AV under test, a failure could be a collision, driving dangerously close to other vehicles or obstacles, or being unable to drive within the lanes etc. 

\subsection{Generation of Instance Space}

\begin{figure}[!ht]
    \centering
    \includegraphics[width=\linewidth]{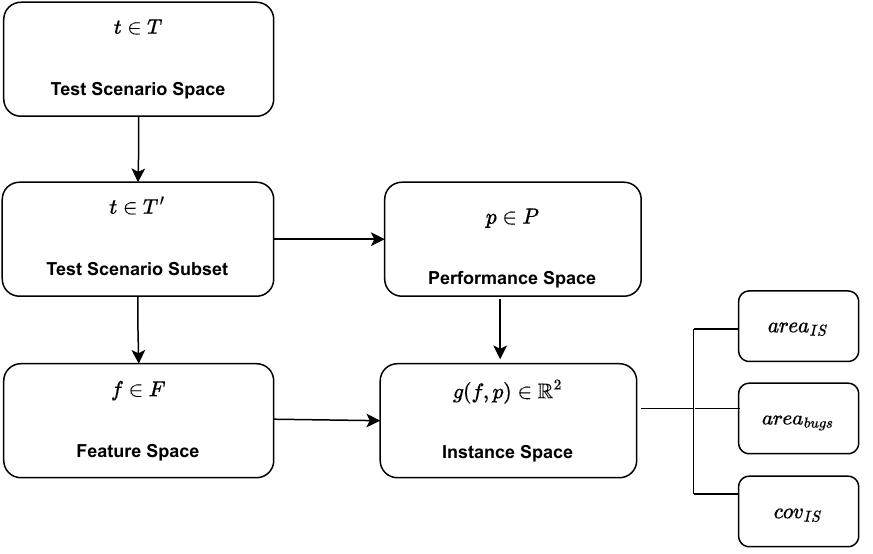}
    \caption{Instance Space Analysis for AV testing.}
    \label{fig:ISA}
\end{figure}

For the generation of instance space, three types of spaces are required (Figure~\ref{fig:ISA}):

\begin{itemize}
    \item Test Scenario Space $T$: In the context of testing AVs, $T$ encompasses all potential test scenarios that could be utilised. Within this space, a specific subset of scenarios, labelled as $T'$ is chosen to comprise the test suite used in the present study.
    \item Feature Space $F$: It consists of a vector of meaningful features that define a test scenario. Features are domain-specific, and their extraction requires significant domain knowledge~\cite{munoz2018instance,mario2020regression}.
    \item Performance Space $P$: It represents the performance of the test scenarios measured on the basis of the metric to assess the effectiveness of a scenario, e.g., number of collisions, violation of safety distance, out-of-bound episodes (OBEs) etc.
\end{itemize}

An instance space is the $2D$ representation of the test scenarios, defined in terms of features having maximum impact on scenario outcome. Therefore, \textit{identification of the features having maximum impact on test outcome} is one of the most crucial steps in ISA. Feature identification and selection is an iterative process that uses machine learning techniques to find significant features that clearly differentiate effective, i.e., failure-revealing scenarios, from the passing ones. The initial stage of this process involves identifying a cluster of features that share similarities with each other. We use k-means clustering for this process as it is one of the simplest unsupervised machine learning algorithms and ISA is shown to perform well with this clustering technique in the previous studies~\cite{smith2022instance, munoz2017performance,munoz2018instance,mario2020regression}. The optimal number of clusters to be used by k-means is selected using silhouette analysis~\cite{aranganayagi2007clustering}. \color{black}

Once clusters are created, one feature from each cluster is taken to create a feature set. Assume that the number of clusters generated in the previous step is $n$, then each feature set will contain $n$ features. This n-dimensional feature set is projected to a temporary $2D$ space using \textit{Principal Component Analysis}~\cite{abdi2010principal}. The process is repeated for all the feature sets created by the possible combination of features from all the clusters. The coordinates of the temporary $2D$ spaces then become the input to a set of Random Forest (RF) models, which learn the feature combination giving the lowest predictive error in predicting the scenario outcome.

Now that the set of most effective features has been identified, we project the test instances defined in terms of $nD$ feature space to a $2D$ coordinate system in such a way that the relationship between the features of the instances and the test outcomes can easily be identified. An ideal projection is one that creates a linear trend when each feature value and scenario outcome is inspected, i.e., low values of features/scenario outcomes at one end of a straight line and high values at the other. Furthermore, the instances that are neighbours in high dimensional feature space should remain as neighbours in the 2D instance space (topological preservation). These projection goals are achieved by using a projection method called \textit{Projecting Instances with Linearly Observable Trends} (PILOT)~\cite{munoz2018instance}, which seeks to fit a linear model for each feature and test outcome, based on the instance location in the $2D$ plane. Mathematically, this involves solving the following optimisation problem:

%
\begin{eqnarray}
	\min					&& \left\| \Ft - \B_{r}\Z \right\|^{2}_{F} + \left\| \Y - \Cbf_{r}\Z \right\|^{2}_{F} 
	\label{eq:optimisation}
	\\
	\text{s.t.}				&& \Z = \A_{r}\Ft \label{eq:score}
	\nonumber
\end{eqnarray}

\noindent where $\Ft$ is the matrix containing the $n$ features of a test scenario, $\Y$ is the column vector containing scenario outcome, $\Z\in\Rbb^{i\times 2}$ is the matrix containing $z_1$ and $z_2$ coordinate values of $i$ scenarios in the $2D$ space, $\A_{r} \in \Rbb^{2 \times n}$ is a matrix that takes the feature values and projects them in $2D$ space, $\B_{r} \in \Rbb^{n \times 2}$ is a matrix that takes the $2D$ coordinates and produces an estimation of the feature values, and $\Cbf_{r}\in\Rbb^{t\times 2}$ is a matrix that takes the $2D$ coordinates and makes an estimation of the technique's performance. In short, Equation~\ref{eq:optimisation} finds the difference between the actual values of the features and performances in a higher dimension and the estimation of these values in $2D$. The lower the difference, the higher the topological preservation. Mathematical proofs and additional technical details of PILOT are available in ~\cite{munoz2018instance}. \color{black}
\subsection{Computation of TISA Metrics}
\label{sec:ISA-metrics}
\begin{figure*}[!t]
    \subfloat[Scenario outcome distribution]{\includegraphics[width=0.24\linewidth]{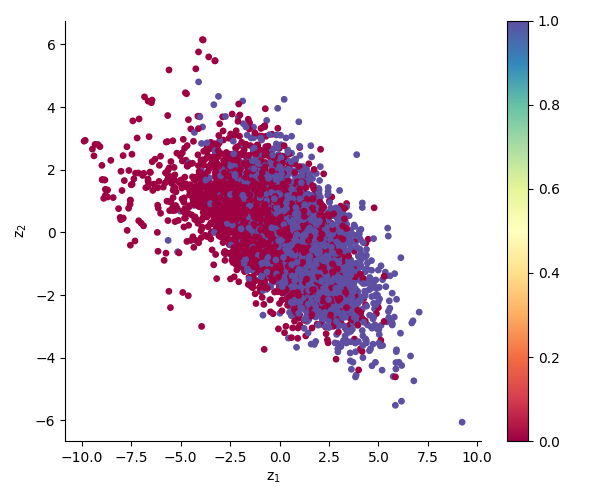}
    \label{fig:performance_distribution}}%
	\subfloat[Num right turns]{\includegraphics[width=0.24\linewidth]{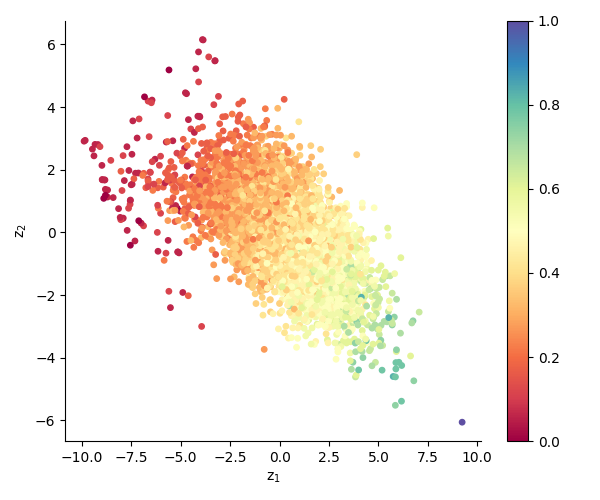}\label{fig:dist_feat_r_turns}}%
    \subfloat[total angle]{\includegraphics[width=0.24\linewidth]{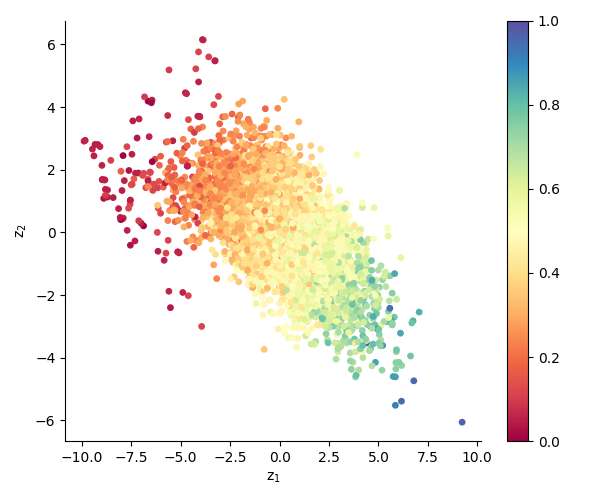}\label{fig:dist_feat_total_angle}}%
\subfloat[footprint]{\includegraphics[width=0.24\linewidth]{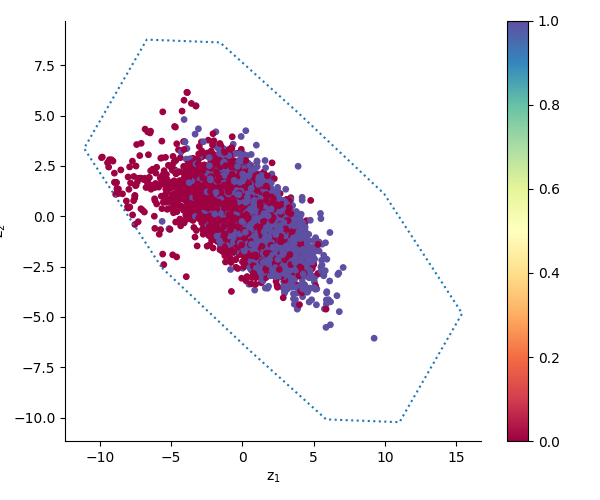}\label{fig:boundary}}%
	\caption{Instance Space Analysis for AV testing (a) shows the distribution of scenarios' outcome; (b and c) show the feature distributions that can be mapped to the scenario outcome; (c) shows the mathematical boundary created around the instance space for estimating the coverage of the test suite. }
	\label{fig:ISA_AV}
\end{figure*}

The adequacy metrics proposed by ISA can not be understood completely without visualising the instance space itself. For this, using the methodology explained in the above sections, we create an instance space and present its important components in Figure~\ref{fig:ISA_AV}. The test suite for the generation of this instance space is obtained from a previous study on test-case prioritisation for AVs~\cite{birchler2021automated}. Detailed information about the features of these test scenarios is provided in Section~\ref{sec:dataset}, where we discuss the dataset and other relevant aspects of our experiments.

Figure~\ref{fig:performance_distribution} shows the distribution of test scenarios in the instance space based on their outcome. The blue points represent the effective test scenarios that push the AV to a safety critical situation (failed), while the red points represent safe scenarios (pass). It can be seen that the scenarios are distributed in such a way that there is a clear distinction between safe and unsafe scenarios, and it is easy to visualise the diversity of the bugs, as well as the diversity of the whole test suite. Figures~\ref{fig:dist_feat_r_turns} and~\ref{fig:dist_feat_total_angle} show the distribution of two of the features of test scenarios across the instance space. These are the features of the test instances that are identified to have the maximum impact on scenario outcome. The feature distributions are projected in the instance space in such a way that feature values are increasing/decreasing from one end of the space to the other, and are easy to map to the test outcome. It can be seen that for both features, the lower values result in safer scenarios, while medium to high values cause failure. 

\subsubsection{Area of the Instance Space ($area_{IS}$)}\textit{This is the area occupied by all the test instances enclosed by the instance space and measures the diversity of the whole test suite}. In Figure~\ref{fig:performance_distribution}, $area_{IS}$ represents the area of the region occupied by all the red and blue dots. 

$area_{IS}$ provides a robust and defensible test adequacy metric, as increasing this area results in increasing the coverage of a broad range of test conditions and, therefore, improves the quality of testing. It is important to note here that the value of $area_{IS}$ will not increase by adding more instances to the instance space, until the feature values of new instances are different from the previous ones and add more diversity to the test suite. Test scenarios having similar features project on top of or very close to each other, therefore, do not increase the area of the instance space. 

\subsubsection{Area of Buggy region ($area_{bugs}$)}\textit{This is the area of the instance space occupied by the majority of the test instances that reveal failure}. In Figure~\ref{fig:performance_distribution}, this is the area of the instance space dominated by blue points (right half of the instance space). 

The number of bugs revealed by a test suite is a common measure of assessing the quality of testing for an AI-based system~\cite{tian2022mosat, li2020av, almanee2021scenorita, abdessalem2018testing}. However, for any AI-based system in general, and for AVs in particular, solely focusing on finding more bugs may overlook critical safety issues if the test suite does not cover a diverse range of scenarios, environmental conditions, and edge cases. Therefore, the above-proposed measure of assessing the quality of a test suite in terms of the diversity of the bugs detected appears to be both reasonable and robust.

\textbf{Area calculation:}
For area calculation, we rely only on high-density regions of instance space to ensure that the values of $area_{IS}$ and $area_{bugs}$ won't get impacted by the outliers. For this, we use DBSCAN~\cite{schubert2017dbscan}, a clustering algorithm that defines a cluster as a dense region of points. DBSCAN takes two parameters $\left\{k,\varepsilon\right\}$ as input, where $k$ represents the minimum number of instances required to form a dense region, while $\varepsilon$ represents the maximum radius of an instance required for it to be considered as a neighbour. The metric used to measure the distance is Euclidean. The values for $k$ and $\varepsilon$ are chosen automatically as recommended in~\cite{daszykowski2001looking}, using Equations~\ref{eq_DBSCAN_params1} and~\ref{eq_DBSCAN_params2}.

\begin{equation}
    k \leftarrow \max\left(\min\left(\left\lceil r/20\right\rceil,50\right),3\right)
    \label{eq_DBSCAN_params1}
\end{equation}

\begin{equation}
    \varepsilon \leftarrow \frac{k\Gamma\left(2\right)}{\sqrt{r\pi}}
    \left(\text{range}\left(z_{1}\right)\times\text{range}\left(z_{2}\right)\right)
    \label{eq_DBSCAN_params2}
\end{equation}

Where $r$ is the number of failed scenarios in the calculation of $area_{bugs}$ and all the scenarios otherwise, $\Gamma\left(\cdot\right)$ is the Gamma function, and ${z_1, z_2}$ are the coordinates of the $2D$ instance space. 

The footprint of a cluster is created using an $\alpha$-shape, which is a generalisation of the convex hull concept from computational geometry~\cite{edelsbrunner1983shape}. It corresponds to a polygon that closely surrounds all the points within a cloud. An $\alpha$-shape is created for each cluster,
and all shapes are bounded together as a MATLAB polygon structure~\cite{polygons}. The area of the polygon is then computed using MATLAB Polyarea function~\cite{Polyarea}.

In calculating the area of the bugs ($area_{bugs}$), there could be regions of the space where there is an overlap of buggy and bug-free test scenarios. The decision of whether to include such a region in the area calculation depends on the relative number of buggy instances in this area. If the number of bugs is higher than the bug-free instances, the region will be included, ignored otherwise.

\subsubsection{Coverage of Instance Space ($cov_{IS}$)}
The instance space is the $2D$ representation of the test instances available in the test suite under study. These are represented as \textit{Test Scenario Subset} ($T'$) in Figure~\ref{fig:ISA}. However, these scenarios do not necessarily include all the test instances required to explore the feature space completely. Using ISA, we devise a way to compute a mathematical boundary that encloses all the test scenarios which are empirically possible to generate, though, may be missing from the current instance space/test suite. This complete set of scenarios is indicated as~\textit{Test Scenario Space} ($T$) in Figure~\ref{fig:ISA}. 

\textbf{Boundary Creation}: Let $\mathbb{R}^{n\times n}$ be the correlation matrix of $n$ features, defining a test scenario. We define two vectors  $\mathbf{f}_U = \begin{bmatrix} f_{U_1} \cdots f_{U_n} \end{bmatrix}^T$ and $\mathbf{f}_L = \begin{bmatrix}f_{L_1} \cdots f_{L_n}\end{bmatrix}^T$
containing the upper and lower bounds of feature values. We define a vertex vector from these two vectors containing a combination of values from $f_U$ and $f_L$ such that only the upper or lower bound of a feature is included. For instance, $\mathbf{v}_1 = \begin{bmatrix}f_{U_1} f_{L_2} \cdots f_{L_n}\end{bmatrix}^T$ represents a vertex vector containing the maximum value of feature 1 and minimum values of all the other features. We define a matrix $\mathbf{V} = \begin{bmatrix}\mathbf{v}_1 \cdots \mathbf{v}_q\end{bmatrix} \in \mathbb{R}^{n\times q}, q = 2^n$ containing all possible vertices created by the feature combinations. The vertices in metric $\mathbf{V}$, connected by edges, define a hyper-cube that surrounds all the instances in the instance space. 

Some of the vectors in $\mathbf{V}$ represent feature combinations that are unlikely to coexist. For example, if features 1 and 2 are strongly positively correlated, an instance having a high value of feature 1 and a low value of feature 2 is unexpected to be found. Therefore, a vertex vector $\mathbf{v} = \begin{bmatrix}f_{U_1} f_{L_2} \cdots f_{L_n}\end{bmatrix}^T$ would be unlikely to be near any true instance. Similarly, a vertex vector cannot simultaneously contain $\{f_{U_1}, f_{U_2} \}$ or $\{f_{L_1}, f_{L_2} \}$ if feature 1 and feature 2 are strongly negatively correlated. All such unlikely vertex vectors are eliminated from $\mathbf{V}$. The edges connecting the remaining vertex vectors are then projected into $2D$ instance space using PILOT~\cite{munoz2018instance}, whose convex hull now represents the mathematical boundary of the expanded instance space. Figure~\ref{fig:boundary} shows the boundary created around the instance space shown in figure~\ref{fig:performance_distribution}.

The coverage of the instance space is defined as \textit{the percentage of the bounded area ($area_{bound}$) covered by the instance space}. 
\begin{equation}
\label{eq:coverage}
\texttt{Cov}_{IS} (\%) = \frac{\texttt{area}_{IS}}{\texttt{area}_{bound}} \, 100 
\end{equation}

As can be seen in figure~\ref{fig:boundary}, there are vast empty regions around the instance space, depicting that the test suite lack sufficiency to test the system reliably. Increasing the coverage of the space, especially in the \textit{high bug probability area}, would improve the effectiveness of the test suite by testing a wider range of conditions and scenarios. 

\section{Experimental Design}\label{sec:design}

\subsection{Research Questions}
Our empirical study is steered by the following research questions:

\subsubsection{RQ1. How effective are the TISA metrics?} 
Under this research question, we want to investigate the effectiveness of ISA metrics in terms of fault detection. This is done by performing the correlation analysis of the TISA metrics computed for the samples of test scenarios with the number of bugs detected ~\cite{feldt2016test, aghababaeyan2021black, chen2020deep}. In our experiments, the buggy scenarios are the ones that lead AV to collision, break safety distance or cause out-of-bound episodes (OBE). These are the widely used measure of failure in the system testing of AVs~\cite{lu2021search, birchler2022cost,li2020av, piazzoni2021vista, almanee2021scenorita,gambi2019automatically}. We also investigate such correlation for the other diversity metrics selected for this study. 

\subsubsection{RQ2. How do the TISA metrics correlate with existing diversity adequacy metrics?}

There is a plethora of literature available on diversity measures for testing~\cite{feldt2016test, feldt2008searching}. However, different diversity metrics are designed to measure the diversity of the underlying dataset from a different perspective. For instance, Euclidean distance is a measure of the straight-line distance between two points in a multidimensional space. It is commonly used to measure the similarity or dissimilarity between two test scenarios based on the values of their features. On the other hand, the Shannon diversity index is a measure of the diversity of a set of objects based on their relative abundance. It is commonly used to measure the diversity of test scenarios in terms of their frequency or distribution across different categories or features. Both measures can be useful in assessing test diversity, however, their effectiveness is dependent on the specific context and goals of the testing process. Under this research question, we investigate how TISA metrics correlate with other state-of-the-art diversity metrics to investigate their uniqueness or redundancy.  

\subsubsection{RQ3. How efficient are the ISA adequacy metrics?}
We want to compare ISA metrics with other diversity metrics in terms of computation time. As testing of sophisticated AI systems, like AVs, requires executing an enormous number of test scenarios to cover wide driving conditions, computation time becomes one of the leading factors in deciding the acceptability of a metric. 

\begin{figure*}[!t]
    \subfloat[$area_{IS}$]{\includegraphics[width=0.14\linewidth]{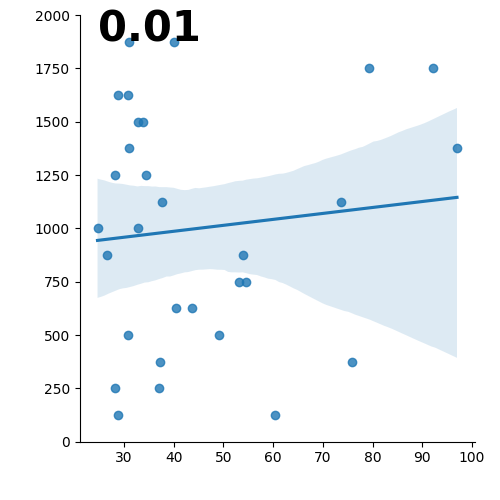}\label{fig:avg_spl}}%
	\subfloat[$cov_{IS}$]{\includegraphics[width=0.14\linewidth]{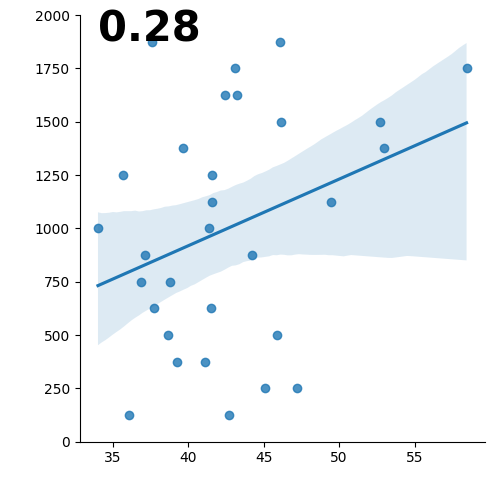}\label{fig:std_cc}}%
	\subfloat[$area_{bugs}$]{\includegraphics[width=0.14\linewidth]{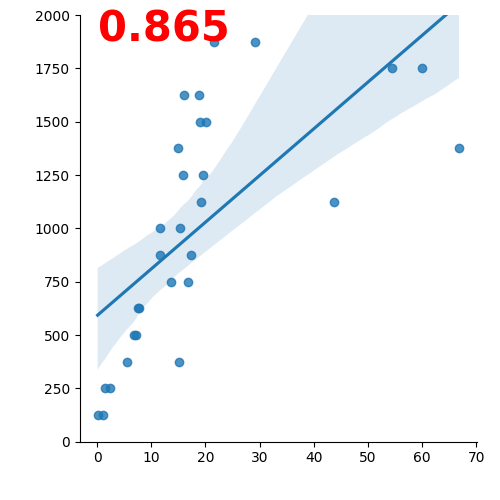}\label{fig:loc}}%
	\subfloat[$div_{Eu}$]{\includegraphics[width=0.14\linewidth]{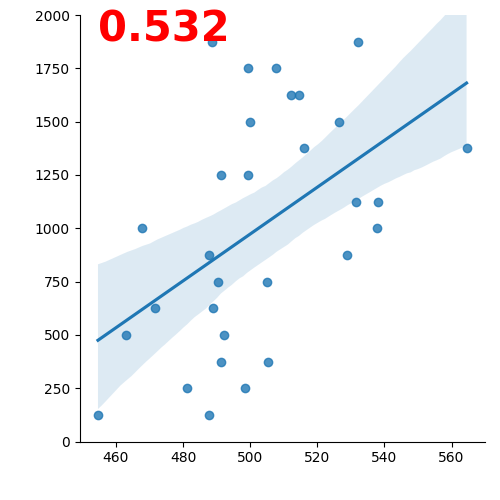}\label{fig:num_tc}}%
	\subfloat[$div_{Shannon}$]{\includegraphics[width=0.14\linewidth]{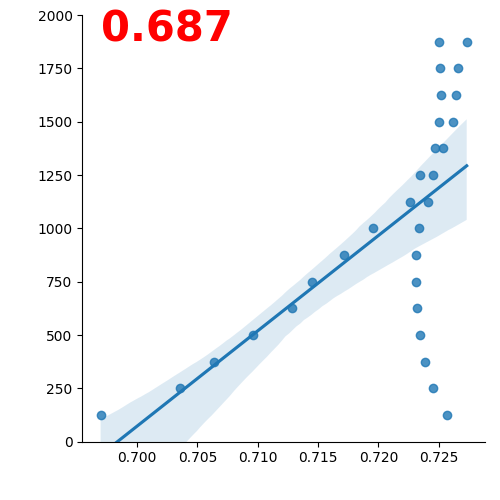}\label{fig:per_cc10}}%
	\subfloat[$NCD$]{\includegraphics[width=0.14\linewidth]{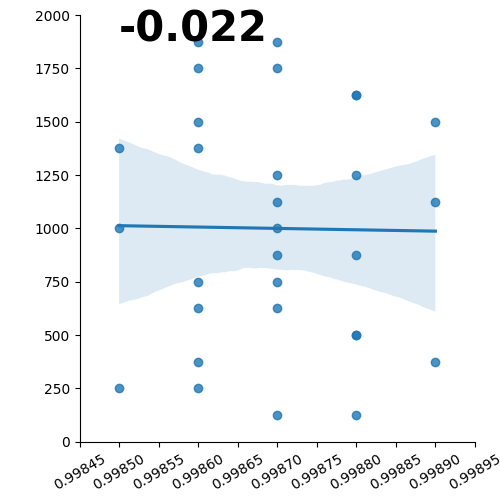}\label{fig:avg_cc}}
	\subfloat[$STD$]{\includegraphics[width=0.14\linewidth]{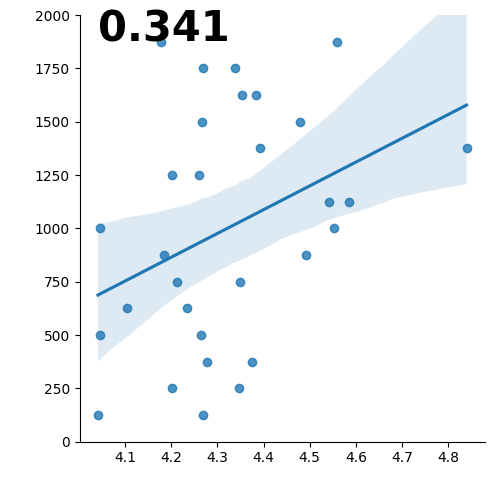}\label{fig:avg_rad}}\\%
	\caption{Correlation between selected diversity metrics and number of bugs Using Spearman's Rank Correlation Coefficient: test suite 1}
    \label{fig:correlation_div_spectre}
\end{figure*}

\begin{figure*}[!t]
    \subfloat[$area_{IS}$]{\includegraphics[width=0.14\linewidth]{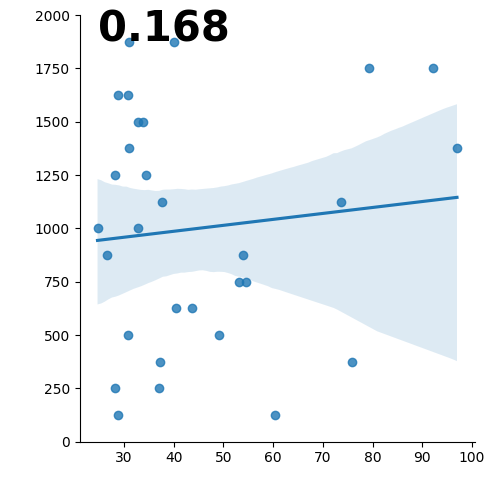}\label{fig:avg_spl}}%
	\subfloat[$cov_{IS}$]{\includegraphics[width=0.14\linewidth]{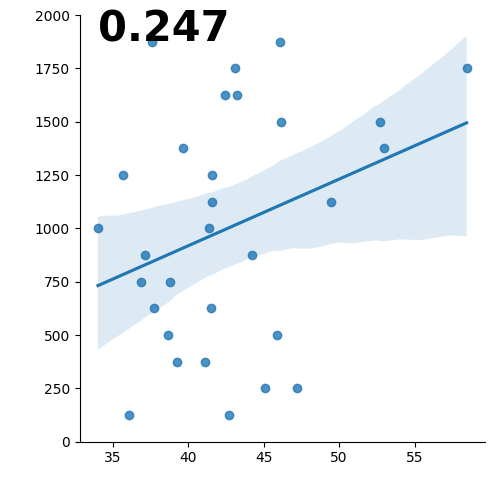}\label{fig:std_cc}}%
	\subfloat[$area_{bugs}$]{\includegraphics[width=0.14\linewidth]{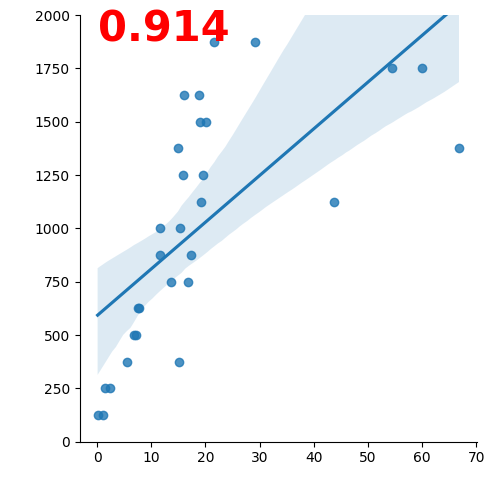}\label{fig:loc}}%
	\subfloat[$div_{Eu}$]{\includegraphics[width=0.14\linewidth]{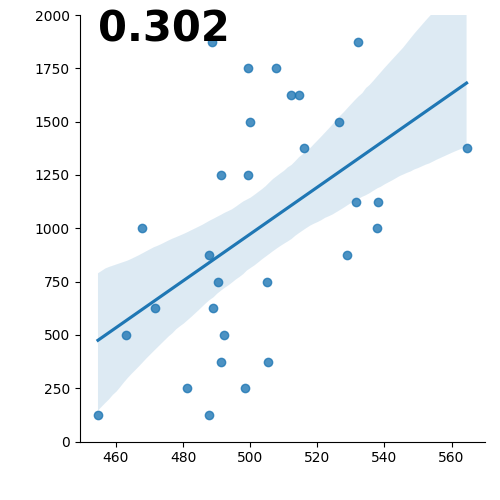}\label{fig:num_tc}}%
	\subfloat[$div_{Shannon}$]{\includegraphics[width=0.14\linewidth]{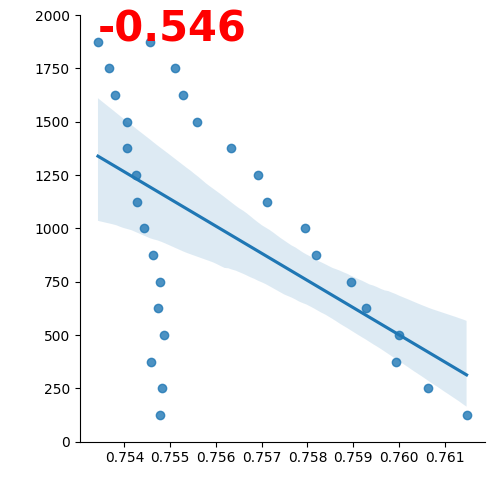}\label{fig:per_cc10}}%
	\subfloat[$NCD$]{\includegraphics[width=0.14\linewidth]{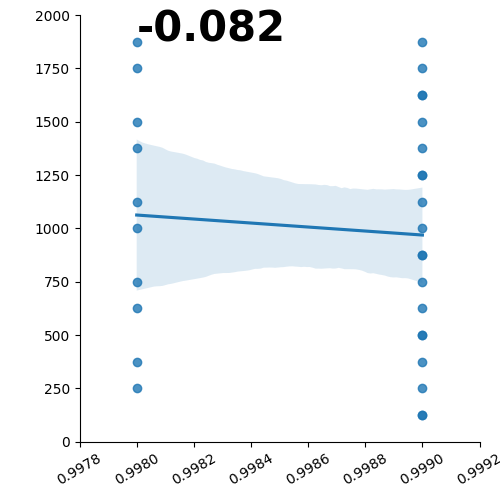}\label{fig:avg_cc}}
	\subfloat[$STD$]{\includegraphics[width=0.14\linewidth]{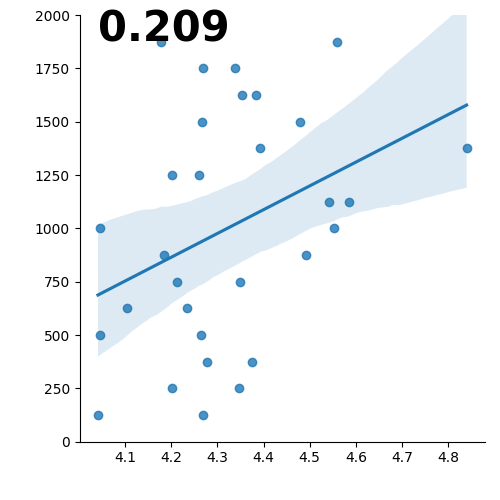}\label{fig:avg_rad}}%
	\caption{Correlation between selected diversity metrics and number of bugs Using Spearman's Rank Correlation Coefficient: test suite 2}
    \label{fig:correlation_div_sdc}
\end{figure*}
\subsection{Dataset}
\label{sec:dataset}
The current study employs test suites from two previous studies on search-based test selection and prioritisation for AVs~\cite{lu2022learning,birchler2021automated}.

The first test suite is generated for testing Baidu Apollo Open Platform 5.0 using the LGSVL simulator~\cite{rong2020lgsvl}. It includes scenarios representing different types of roads in San Francisco. Failures in this dataset are defined as collisions or safety distance violations~\cite{shalev2017formal}. The suite consists of 28,964 unique test scenarios, with 12,932 (45\%) being buggy scenarios.

The second test suite, generated by As-Fault~\cite{gambi2019asfault, gambi2019automatically}, combines search-based testing and procedural content generation for lane-keeping feature testing. Failures in this dataset are defined as Out Of Bound Episodes (OBEs). The scenarios focus on road features like turns and straight segments. The suite contains 5,639 test scenarios, with 2,541 (45\%) being buggy scenarios.

\subsection{Sampling for Experiments}
\label{sec:sampling}

To conduct a correlation analysis between adequacy metrics and the occurrence of bugs, it is crucial to have a diverse range of test suite samples with varying bug counts. To achieve this, we follow a systematic approach of generating samples by modifying the number of bugs within a range of 5\% to 75\%.

To begin, we divide the dataset into two subsets: one comprising failed scenarios and the other consisting of safe driving scenarios. From the failed test scenarios, we sample $n$\% of the scenarios, while the remaining $100 - n$\% is sampled from the safe subset. By combining these sampled scenarios from both subsets, we create an experimental sample.

In total, we generate 30 such samples for our experiments, each containing 2500 scenarios. We start with $n = 5$ and increment it by 5 for each subsequent sample until $n$ reaches 75. If $n$ reaches 75 and the number of samples is still less than 30, we reset the value of $n$ to 5 and continue generating additional samples.
\color{black}
\section{Experimental Results}\label{sec:results}

\subsection{RQ1. Effectiveness of TISA metrics}
\label{sec:result-RQ1}
We aim to investigate the effectiveness of TISA metrics in terms of how well they correlate with fault detection. 

For correlation analysis, we use Spearman Rank Correlation, as it is a non-parametric method and does not require assumptions of linearity or normality of data~\cite{de2016comparing}, therefore, is widely used in similar studies for correlation analysis~\cite{zhang2021fairness, aghababaeyan2021black, dong2019there, dong2020empirical}. \color{black}

Figures~\ref{fig:correlation_div_spectre} and~\ref{fig:correlation_div_sdc} show the correlation of TISA and other selected metrics with fault detection. The x-axis plots the diversity values computed by different diversity measures, while the count of bugs is shown at y-axis. The correlation value ($\rho$) is shown in red if the correlation is statistically significant (p-value $<= 0.05$), black otherwise. 

To consider a measure as a good estimator of fault detection, we anticipate a high and statistically significant correlation. This correlation should be consistent across different test suites. Our observations indicate that all TISA metrics exhibit a positive correlation with the number of bugs in both test suites. Among these metrics, the $area_{bugs}$ demonstrates a very high correlation that is statistically significant. On the other hand, the correlation for the $cov_{IS}$ metric is moderate and statistically insignificant. As for the $area_{IS}$ metric, it shows a very low correlation with bugs in both test suites. 

Euclidean Distance ($div_{Eu}$) shows a moderate to high positive correlation with bugs in our experiments. However, this correlation is not consistently significant across the datasets.

$div_{Eu}$ and $area_{bugs}$  measure the spatial separation between data points in a multidimensional and $2D$ space, respectively. As the bug occurrences in the test suites under study exhibit distinct spatial patterns or clusters (as shown in Figure~\ref{fig:performance_distribution}), such metrics may better capture the proximity or dissimilarity between data points, thereby reflecting the presence of bugs more accurately as compared to other metrics. As the instances in $2D$ instance space are projected in a manner that aligns the distribution of features with the distribution of test outcomes, $area_{bugs}$, which is based on the instance space, exhibits a stronger correlation with the distribution of bugs compared to the $div_{Eu}$.

Among other metrics, $div_{Shannon}$ shows a strong positive correlation in test suite 1, however, the relationship becomes negative in test suite 2. Due to the correlation value of this metric changing signs across the test suite, its effectiveness is questionable, and we cannot place confidence in these findings. $div_{Shannon}$ focuses on the richness and evenness of test instances. It can help identify datasets with a high richness of different variables or dimensions, and can also highlight datasets where the abundance is distributed more evenly or unevenly across the variables. However, it may not capture other aspects of the data's diversity, such as patterns, interactions, or specific relationships between variables. Therefore, this measure doesn't show a consistent relationship with the number of bugs in our experiments. 

Both $NCD$ and $STD$ exhibit either no correlation or statistically insignificant correlations with bugs in both test suites. $NCD$ is primarily designed to measure the similarity between two data sequences based on their compressed lengths. However, in our test suites, where the majority of input features are numeric (e.g., speed, distance, throttle, number of right turns, total angle), it is challenging to represent these numeric variables as data sequences that can be effectively compressed. As a result, $NCD$ does not serve as a strong measure of diversity in our experiment.

On the other hand, $STD$ is sensitive to outliers, which can disproportionately affect its calculation and potentially lead to misleading interpretations of data spread or diversity. Moreover, $STD$ assumes a symmetric distribution around the mean, which may not accurately represent the dispersion of values in the presence of skewed distributions.

These limitations in the effectiveness of $NCD$ and $STD$ emphasise their limitations in accurately capturing diversity in our study.
\color{black}

\begin{custombox}[Answer to RQ1] 
All TISA metrics exhibit a positive correlation with faults, but $area_{bug}$ stands out with a consistently strong and statistically significant correlation, highlighting its effectiveness in detecting faults and reliability in providing valuable insights into the effectiveness of the testing process.
\end{custombox}

\subsection{RQ2. Relationship of TISA metrics with selected diversity metrics}
\label{sec:result-RQ2}

We want to investigate how TISA metrics correlate with other adequacy metrics selected for this study. Understanding the correlation between test adequacy metrics can provide useful insights into redundancy or overlaps between them. For instance, if two metrics are highly correlated, they are likely to measure the same aspect of diversity, and thus it may be more efficient to use only one of them. 

Section~\ref{sec:result-RQ1} shows that $area_{IS}$ has a weak correlation with the bugs in the system, while this correlation is not statistically significant for $cov_{ISA}$. As we are interested in determining the effectiveness of TISA measures in terms of fault detection, we will consider only $area_{bugs}$ for answering RQ2 and RQ3. 

Figures~\ref{fig:correlation_div_spectre} and~\ref{fig:correlation_div_sdc} show the correlation of $area_{bugs}$ with $div_{Eu}$, $div_{Shannon}$, $NCD$ and $STD$ for test suites 1 and 2. The correlation values are coloured red if the correlation is statistically significant, and black otherwise. The metric that shows the strongest positive correlation with $area_{bug}$ is $div_{Eu}$, however, this correlation is not consistently significant across the test suites. $STD$ shows a similar trend as $div_{Eu}$; positively strong and statistically significant correlation for one test suite, however, weak and insignificant correlation for the other. Just like the correlation with the number of bugs, $div_{Shannon}$ changes the sign of correlation across test suites, therefore making the results unreliable. Lastly, $NCD$ consistently shows a negative and weak correlation with $area_{bugs}$.  

\begin{custombox}[Answer to RQ2]
In conclusion, the TISA metric, $area_{bug}$, exhibits a weak or inconsistent correlation with other diversity metrics, suggesting that it captures data diversity from a unique perspective compared to the other metrics analysed. However, due to its strong correlation with the number of bugs and its emphasis on high-impact features, $area_{bug}$ is suitable for use as a black-box adequacy measure for diversity.
\end{custombox}

\begin{figure}[ht]
    \subfloat[$div_{Eu}$]{\includegraphics[width=0.33\columnwidth]{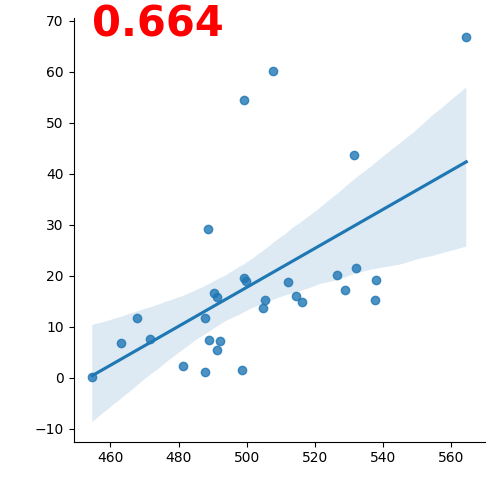}\label{fig:avg_spl}}%
	\subfloat[$div_{Shannon}$]{\includegraphics[width=0.33\columnwidth]{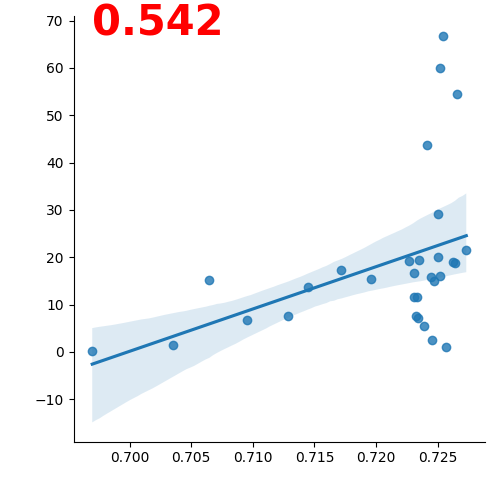}\label{fig:loc}}\\%
	\subfloat[$NCD$]{\includegraphics[width=0.33\columnwidth]{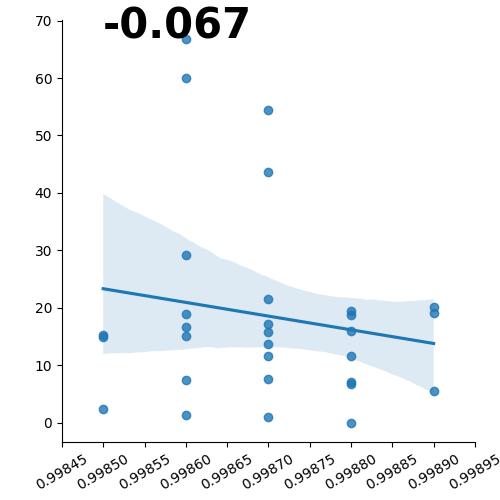}\label{fig:nd}}%
	\subfloat[$STD$]{\includegraphics[width=0.33\columnwidth]{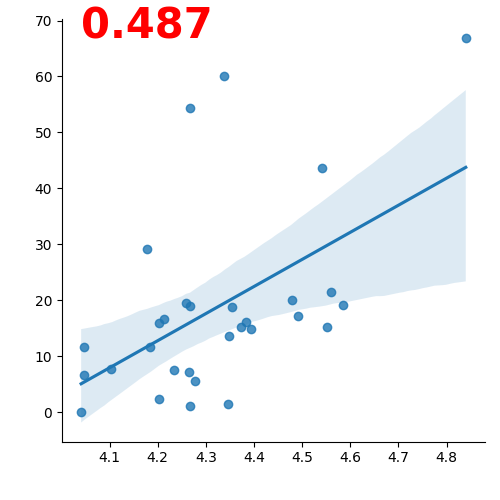}\label{fig:num_tc}}%
	
	\caption{Spearman's Rank Correlation between $area_{bug}$ and other diversity metrics for test suite 1}
    \label{fig:feat_dist}
\end{figure}

\begin{figure}[ht]
    \subfloat[$div_{Eu}$]{\includegraphics[width=0.33\columnwidth]{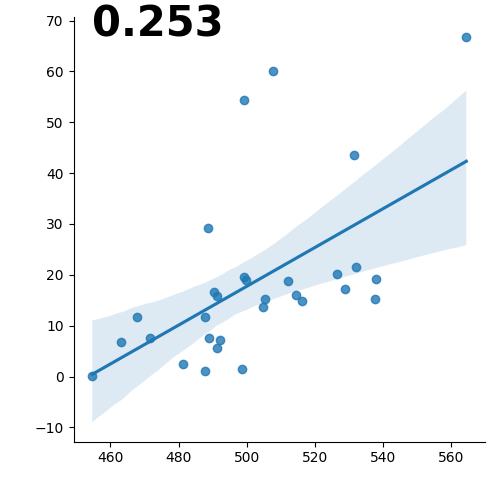}\label{fig:avg_spl}}%
	\subfloat[$div_{Shannon}$]{\includegraphics[width=0.33\columnwidth]{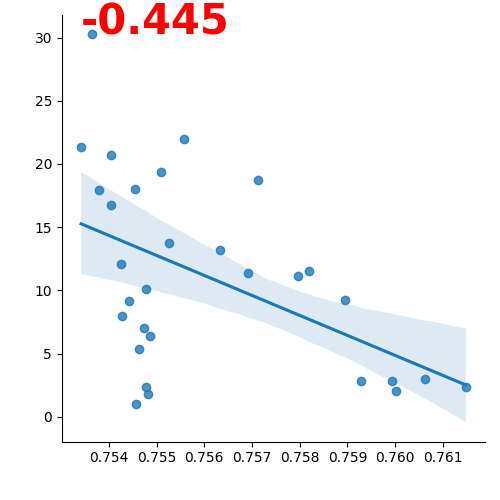}\label{fig:loc}}\\%
	\subfloat[$NCD$]{\includegraphics[width=0.33\columnwidth]{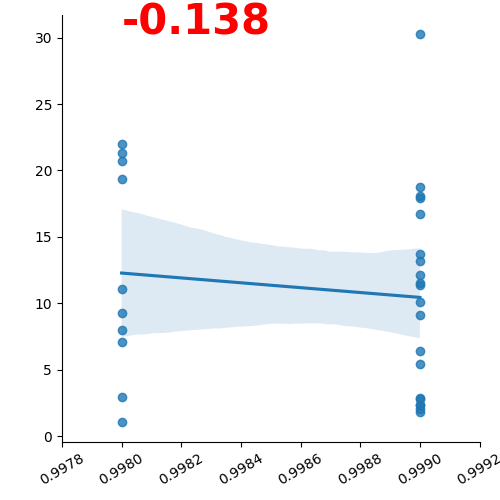}\label{fig:nd}}%
	\subfloat[$STD$]{\includegraphics[width=0.33\columnwidth]{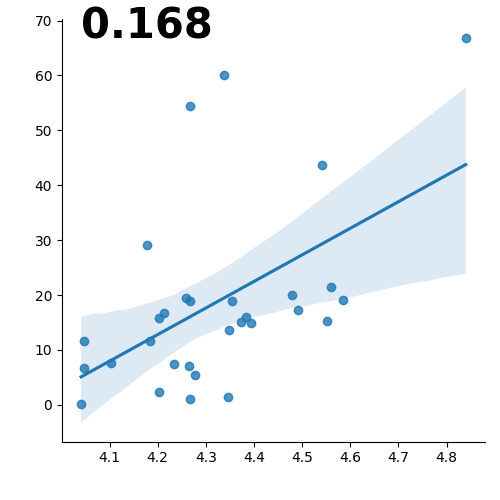}\label{fig:num_tc}}%
	
	\caption{Spearman's Rank Correlation between $area_{bug}$ and other diversity metrics for test suite 2}
    \label{fig:feat_dist}
\end{figure}

\subsection{RQ3. The efficiency of TISA metrics}
\begin{figure*}[ht]
    \subfloat[Execution time of adequacy metrics for different sample sizes]{\includegraphics[scale=0.5]{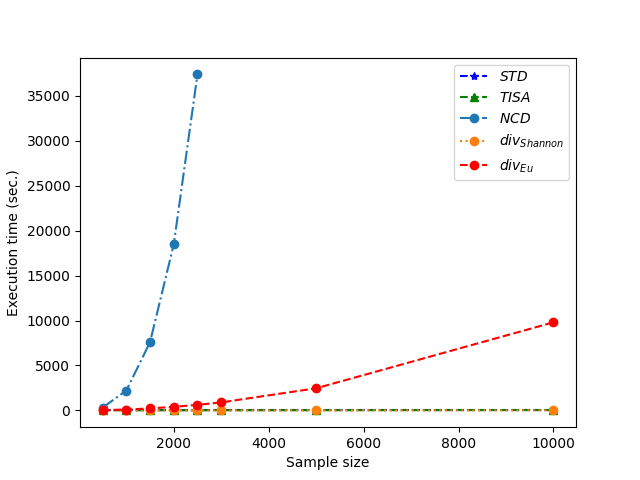}\label{fig:exec_time}}%
    \subfloat[Magnified view of the execution time of $STD$, $TISA$ and $div_{Shannon}$ from Figure~\ref{fig:exec_time} ]{\includegraphics[scale=0.5]{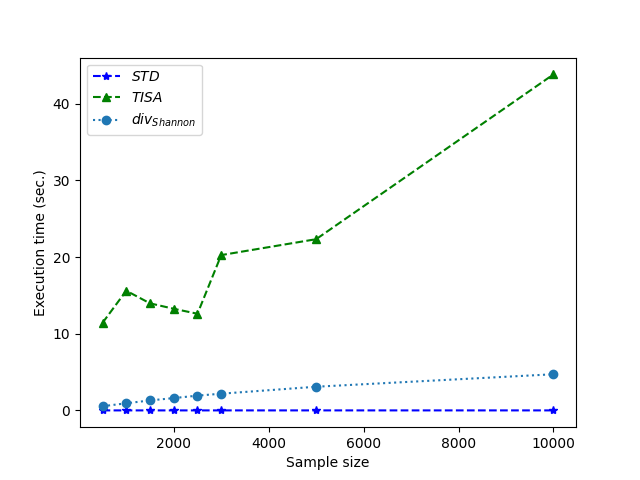}\label{fig:mag_exec_time}}%

	\caption{Computation time taken by adequacy metrics for different sample sizes}
    \label{fig:feat_dist}
\end{figure*}
The testing of AI systems, in particular AVs, requires a huge number of test scenarios to cover wide driving situations on real roads. Therefore, the most appropriate adequacy metric is one that is practically efficient as well as effective. We want to assess the efficiency of the TISA and other selected adequacy metrics in terms of computation time. For this, we selected 8 samples comprising 500, 1000, 1500, 2000, 2500, 3000, 5000 and 10,000 test scenarios. We applied the selected diversity measure to all the samples and recorded the execution time in seconds. Similarly, we generated the instance spaces for all the samples, computed the TISA metrics, and recorded the time of the whole process. All the experiments are conducted on an Ubuntu server with 32 GB memory and an Intel Core 5.20GHz processor. 

The graph presented in Figure~\ref{fig:exec_time} illustrates how the computation time of different metrics changes as the sample size increases. The computation time of $NCD$ is exponentially high compared to all other metrics. The computational complexity of this metric is generally proportional to the length of the input data, and the performance of the compression algorithm. However, a previous study found that even when computing this metric using various compression algorithms to determine the optimal one, the computation time for NCD remains impractical for larger datasets, even with the best compression algorithm~\cite{aghababaeyan2021black}. This makes this metric infeasible for testing AI-based systems, which usually have a huge number of test scenarios, with many features. The execution time of $div_{Eu}$ is less than that of NCD, but it is still considerably higher than other metrics. Moreover, the execution time of $div_{Eu}$ increases exponentially once the sample size surpasses a certain threshold.

To view the trends of $STD$, $TISA$ metrics and $div_{shannon}$ more clearly, which are hidden behind each other in Figure~\ref{fig:exec_time}, we have plotted their computation time separately in Figure~\ref{fig:mag_exec_time}. The computation time of $STD$ remains almost constant, while $div_{Shannon}$ has a slight increase in the time with the increase in sample size. As compared to these two metrics, $TISA$ takes more time to generate the instance space and compute the adequacy results. However, even for a large test suite having 10k test scenarios, this time is still under 50 seconds. It is important to note here that, unlike other metrics, the computation time for TISA includes the generation of instance space, adequacy metrics computation, and generation of feature and performance distribution graphs. Therefore, irrespective of the relative difference of computation time of $area_{bug}$ compared to $div_{Eu}$, $div_{Shannon}$ and $STD$, this metric is quite efficient considering its dual advantage of providing the visual insights into the diversity along with a numeric value. 
\begin{custombox}[Answer to RQ3]
In summary, the computational cost of TISA metrics is not high. Additionally, due to their ability to visualise diversity and effectively detect bugs, these metrics are well-suited for black-box testing of AVs.
\end{custombox}

\section{Related Studies}\label{sec:related-work}
\subsection{Model-level Testing}
The majority of testing methods employed for AVs and other AI-based systems primarily revolve around model-level testing.~\cite{stocco2023model, tian2018deeptest, pei2017deepxplore, zhang2018deeproad, kim2019guiding, deng2020analysis, stocco2022thirdeye, feng2020deepgini, ma2018deepgauge, sun2019structural,  gerasimou2020importance}. 

DeepRoad~\cite{zhang2018deeproad} suggests employing Generative Adversarial Networks (GANs) to generate driving inputs that closely resemble real-world data streams, thereby enhancing the realism of the generated inputs. DeepTest~\cite{tian2018deeptest} leverages affine transformations commonly used in computer vision to generate novel inputs that induce misbehaviour in the deep neural network (DNN). DeepXplore~\cite{pei2017deepxplore} employs white-box testing techniques to generate inputs that aim to achieve maximum coverage of neurons and diverse behavioural outcomes. In their study, Kim et al.~\cite{kim2019guiding} introduce a set of white-box test adequacy criteria based on the notion of surprise. Surprise is defined as the variation in a deep neural network's behaviour when presented with a new test input compared to its training data. The researchers propose generating test inputs that cover a wide range of surprise levels. Deng et al.~\cite{deng2020analysis} examine the robustness of DNN models against adversarial attacks. ThirdEye~\cite{stocco2022thirdeye} utilises attention maps, derived from the field of explainable AI, to anticipate the occurrence of misbehaviour in AVs. Feng et al. propose \textit{DeepGini}~\cite{feng2020deepgini}, a test prioritisation technique based on the principle that a test case for which the model classifies all the labels/classes with similar probabilities indicates that the model has lower confidence in its classification and is more likely to make mistake as compared to a test case with a significantly higher probability for one label compared to others.

In contrast to the aforementioned studies, our research focuses specifically on conducting ``black-box system-level testing'' for autonomous vehicles (AVs). In system-level testing, the DNN is integrated into the operational environment where it is intended to operate, and any identified test failures are characterised in terms of the overall system's malfunction or misbehaviour. Furthermore, unlike the majority of model-based approaches mentioned earlier, our study employs test cases that replicate real driving scenarios through driving simulations. These simulations encompass various elements such as road conditions, weather conditions, AV behaviour, and other vehicles present on the road. In contrast, the datasets used in many of the mentioned model-based approaches typically consist of images as test cases, with the AI model being evaluated primarily for image classification or other tasks at the model level.

\subsection{System-level Testing}
This section reviews recent work done for system-level black-box testing of AVs. Specifically, we'll be discussing the studies using diversity and coverage metrics within this context. 

\subsubsection{Diversity}
Diversity is an apparent indicator of quality because it directly indicates how thoroughly the system is tested across various conditions and scenarios. It is a widely used adequacy measure in traditional software~\cite{aghababaeyan2021black,hauer2019did,arcaini2021targeting,tang2021collision,majzik2019towards,laurent2022parameter} and also deemed important for testing of AI-based systems~\cite{lu2021search, birchler2021automated, ebadi2021efficient, feldt2016test, moghadam2022machine}. We will summarise work related to the use of diversity in black-box testing of AI-based systems in this section. 

Birchler et al. prioritise test cases based on diversity and execution cost using static road features~\cite{birchler2021automated,birchler2022single}. Lu et al. prioritise test cases based on diversity, demand, collision information and collision probability~\cite{lu2021search}. Ebadi et al. generate test scenarios for pedestrian detection, measuring effectiveness based on fault detection and diversity~\cite{ebadi2021efficient}. Moghadam et al. use bio-inspired algorithms to test lane keeping functionality of AVs, measuring effectiveness in terms of diversity and the number of bugs detected~\cite{moghadam2022machine}. Vincenzo et al. use roads' dissimilarity measured in terms of weighted Levenshtein distance to evaluate how two test cases are different from each other~\cite{riccio2020model}. 

What makes our work different from these studies is that we propose a novel set of adequacy metrics that assess the quality of a test suite, both in terms of diversity and coverage. The proposed framework also facilitated visual insights by projecting the test scenarios defined in terms of $n$ features, from $n$-dimensional feature space to $2D$ instances space. 

Aghababaeyan et al. performed a correlation analysis of black-box diversity measures (Geometric Diversity~\cite{kulesza2012determinantal}, Normalised Compression Distance~\cite{cohen2014normalized}, and Standard Deviation) with fault detection~\cite{aghababaeyan2021black}. They found that Geometric Diversity had the highest and statistically significant correlation. Our work is similar in analysing diversity metrics' correlation with bugs but introduces new metrics with medium to high fault detection correlation, outperforming other metrics in fault detection and computation time. Furthermore, Geometric Diversity, assessed on image datasets using VGG-16~\cite{mousser2019deep}, requires a large number of features, which are not feasible for system testing. Therefore, this diversity measure cannot be used in the context of black-box system testing. 

\subsubsection{Coverage}
The majority of the approaches proposing black-box coverage metrics for testing AVs are in the white-box domain and at model level~\cite{pei2017deepxplore, ma2018deepgauge, sun2019structural, kim2019guiding, gerasimou2020importance}. A limited work introducing black-box coverage criteria for testing AI-based systems is discussed below:

Hauer et al.~\cite{hauer2019did} propose a statistical approach to ensure all possible scenario types, like overtaking, lane change, etc., are covered in the test suite. Arcaini et al. consider low-level driving characteristics as scenario types for coverage measurement~\cite{arcaini2021targeting}. Tang et al. classify scenarios based on map topological structures and evaluate coverage accordingly~\cite{tang2021collision}. Laurent et al.~\cite{laurent2022parameter} use weight coverage to cover different configurations of a path planner. Unlike these studies, our work introduces a coverage metric that provides a quantitative measure of quality. Furthermore, our proposed framework is generic and can be applied to the black-box testing of any AI-based system. 

A closely related work is presented by Zohdinasab et al.~\cite{zohdinasab2021deephyperion}, who propose DeepHyperion, a framework that analyses the impact of features on Deep Learning (DL) system quality using illumination search~\cite{mouret2015illuminating}. DeepHyperion evaluates test instances based on a fitness function, visualising the results in a $2D$ map. However, its manual open coding procedure for feature labelling limits its scalability and makes it impractical for testing instances with many features. Additionally, DeepHyperion presents feature maps in an $n$-dimensional coordinate system, which restricts the visualisation of the combined impact of more than three features. In contrast, our framework projects test instances from an n-dimensional feature space to a $2D$ instance space, providing intuitive and easy-to-analyze visualisations.

\section{Threats to Validity}\label{sec:threats}
\textbf{Internal validity threats} arise when the results are influenced by internal factors. One such factor is the lack of source code for the selected diversity metrics, which could affect our implementation. To mitigate this, we extensively tested our code against the original paper. Another threat is the limited variability in the number of bugs in the samples due to random sampling. We addressed this by using a custom sampling method to ensure diverse bug counts. Lastly, the configuration of hyperparameters for machine learning methods could pose a threat. To mitigate this, we employed established techniques such as silhouette analysis~\cite{aranganayagi2007clustering} for k-means and an automated method for parameter calculation for DBSCAN~\cite{daszykowski2001looking}. These measures strengthen the validity of our study.

\textbf{Conclusion threats to validity} pertain to issues that can affect the accuracy and reliability of the study's conclusions. One such threat in this study is the choice of correlation analysis method, which may be influenced by data distributions or the shape of the relationship. To mitigate this, we use Spearman correlation, which does not rely on assumptions about data distributions but requires a monotonic relationship.

Another such threat concerns the validity of feature combinations used for the estimation of the boundary for coverage calculation. These combinations are learned from test scenarios based on feature correlation analysis. If the test suite contains invalid or unrealistic scenarios, incorrect correlations can be learned, rendering the boundary invalid. To address this threat, we utilise test suites from two previous studies that have reported the generation of valid and realistic driving scenarios~\cite{lu2022learning, birchler2021automated}.

\textbf{External validity threats} impact the generalizability of the results. To minimise these impacts, we selected two different test suites having a different number of features and test instances. Furthermore, we have selected the widely used adequacy metrics published in the studies of black-box testing of AI-based systems.

\section{Discussion}

The TISA coverage metric ($area_{IS}$), measures the coverage of a test suite in relation to the entire feature space or extended instance space. It represents the percentage of the area covered by the test suite out of the total possible area. While it may not strongly correlate with bug detection, it serves as a valuable objective measure for evaluating the extent to which the test suite explores the feature space effectively. By visualising the coverage achieved, $area_{IS}$ can offer a graphical representation of the coverage landscape. This visualisation allows the testers to observe the areas that are well covered and those that remain unexplored. It can help identify gaps in testing and guide efforts to improve coverage by targeting specific regions of the feature space.

The accuracy of $area_{IS}$ depends on the feasibility of boundary estimation, which involves determining the upper and lower limits of the features associated with the driving scenarios. In the present study, the boundary calculation utilises the maximum and minimum values of the features that are currently available within the test suite. Consequently, the unoccupied regions in Figure~\ref{fig:boundary} indicate the scenarios that are absent relative to the test suite itself. However, the estimation of absolute feature boundaries by defining Operational Design Domains (ODDs)\cite{ye2022operational} for the feasible driving conditions in which autonomous vehicles (AVs) can theoretically and practically operate, is an area of active research\cite{smart2021autonomous, schwalb2021two}. Some initial efforts have been made to establish these boundaries~\cite{ODDs}. Coverage based on absolute boundaries would provide a measure of how well the entire feature space, as defined by the potential driving conditions, is tested, regardless of the specific test suite employed for generating the instance space.

\section{Conclusion and Future Work}\label{sec:conclusion}
In this study, we introduce a set of adequacy metrics, called $TISA$ metrics, to assess the quality of black-box testing of AI-based systems. The proposed metrics evaluate the test suite both from a coverage and diversity perspective. They also facilitate insights into the diversity of the bugs identified by the test suite by providing the measure of the diversity of the failed test scenarios. Along with giving the measure of diversity in numeric form, the diversity of the test suite is presented in $2D$~\textit{instance space}. The instance space also provides a clear and precise representation of the test coverage, making it easy to identify if tests for a particular feature combination are missing or if there is a need to remove the redundant ones; therefore, it can be a powerful tool to improve the quality of the test suite. 

We assess the effectiveness of the proposed metrics in terms of fault detection. For this, we perform the correlation analysis of these metrics with the number of bugs identified. Furthermore, to test if these metrics are redundant to other diversity metrics common in testing, we performed a correlation analysis of $TISA$ metrics with other metrics. Experimental results show that $TISA$ metrics show a high correlation with bugs, measure a different aspect of diversity as compared to other diversity metrics, and are computationally efficient. 

We expect to extend this work by incorporating TISA metrics in test generation, selection and prioritisation strategies. Furthermore, we plan to generate a test generation strategy that identifies the feature values of the missing scenarios from the instance space and estimates new scenarios using these features. 

\section*{Acknowledgements}

The funding for this research is provided by the Australian Research Council under grant DP210100041.
\bibliographystyle{plain}
\bibliography{references.bib}

\end{document}